*Review*

# Recent Progress in Energy Management of Connected Hybrid Electric Vehicles Using Reinforcement Learning

**Min Hua [1], Bin Shuai [1,2,*], Quan Zhou [1], Jinhai Wang [1], Yinglong He [3], and Hongming Xu [1]**

[1]  School of Engineering, University of Birmingham, Birmingham B152TT, UK
[2]  School of Vehicle and Mobility, Tsinghua University, Beijing 100084, China
[3]  School of Mechanical Engineering Sciences, University of Surrey, Guildford, GU27XH, UK
*  Correspondence: shuaib@mail.tsinghua.edu.cn

**Abstract:** The growing adoption of hybrid electric vehicles (HEVs) presents a transformative opportunity for revolutionizing transportation energy systems. The shift towards electrifying transportation aims to curb environmental concerns related to fossil fuel consumption. This necessitates efficient energy management systems (EMS) to optimize energy efficiency. The evolution of EMS from HEVs to connected hybrid electric vehicles (CHEVs) represent a pivotal shift. For HEVs, EMS now confronts the intricate energy cooperation requirements of CHEVs, necessitating advanced algorithms for route optimization, charging coordination, and load distribution. Challenges persist in both domains, including optimal energy utilization for HEVs, and cooperative eco-driving control (CED) for CHEVs across diverse vehicle types. Reinforcement learning (RL) stands out as a promising tool for addressing these challenges. Specifically, within the realm of CHEVs, the application of multi-agent reinforcement learning (MARL) emerges as a powerful approach for effectively tackling the intricacies of CED control. Despite extensive research, few reviews span from individual vehicles to multi-vehicle scenarios. This review bridges the gap, highlighting challenges, advancements, and potential contributions of RL-based solutions for future sustainable transportation systems.

## 1. Introduction

The rapid expansion of industrialization, urbanization, and technological advancements has ushered in an unprecedented era of heightened global energy consumption. This surge in energy demand not only places strain on existing resources but also raises critical concerns regarding environmental sustainability, largely due to the predominant utilization of fossil fuels [1]. In light of these complex challenges, the electrification of transportation has emerged as a compelling avenue for resolution [1–3]. Consequently, automotive manufacturers are progressively pivoting away from conventional fossil fuel-powered vehicles, embracing innovative energy alternatives such as battery electric vehicles (BEVs), hybrid electric vehicles (HEVs), and fuel cell electric vehicles (FCEVs) [4–6]. Since such electric vehicles (EVs) stand out for their ability to enhance fuel economy, reduce emissions, and extend mileage range while navigating urban and environmental restrictions, however, the main limitations include a limited range compared to HEVs. HEVs allow them to offer the benefits of electrification without the range and charging constraints of BEVs. And FCEVs boast a longer range and faster refueling times compared to BEVs but are limited by the current scarcity of hydrogen refueling infrastructure. However, in response to these challenges, the effective energy management system (EMS) has emerged as a pivotal solution for optimizing energy usage and enhancing efficiency across various sectors.

Further, the development of EMS has undergone a significant transformation due to the transition from HEVs to connected hybrid electric vehicles (CHEVs) [7−9]. Initially, EMS was designed to optimize the energy consumption and efficiency of HEVs, focusing on individual vehicle performance and range enhancement [10,11]. However, as the landscape of electric mobility evolved to include CHEVs such as fleet control, the complexity of managing energy needs escalated. The integration of connectivity features in CHEVs brought new dimensions to EMS, necessitating the evolution of these systems to cater to the unique demands of a connected fleet. These advanced EMS now require sophisticated algorithms for velocity optimization, dynamic charging coordination, and power distribution among multiple vehicles. Additionally, the integration of vehicle-to-vehicle (V2V) and Internet of Things (IoT) technologies became crucial for proactive energy management, allowing for predictive control, enhanced vehicle-to-infrastructure (V2I) interactions, and improvement in overall fleet efficiency. In essence, the background of EMS has evolved from its origins in addressing the energy challenges of HEVs to meeting the intricate energy optimization and distribution posed by the widespread adoption of CHEVs. This evolution marks a pivotal role in shaping the future of sustainable transportation systems, where EMS not only optimizes individual vehicle performance but also contributes to the efficiency and sustainability of broader smart transportation networks [12,13].

The EMS is crucial in HEVs and CHEVs, managing power distribution among various components like the internal combustion engine and battery [14–16]. While there's significant research on EMS, the approaches differ. Rule-based strategies use predefined rules based on the vehicle characteristics and operation modes, adjusting control outputs according to vehicle states like battery state-of-charge and speed. However, these strategies lack flexibility for different driving conditions and require extensive calibration. In contrast, optimization-based methods, including both offline and online options, aim to address these limitations. Offline methods like dynamic programming (DP) use prior knowledge of driving cycles to find optimal policies, but they are tough to apply in real-world conditions, making them less suitable for meeting different vehicle performances, such as comfort, fuel efficiency, and automation. Online optimization methods, such as model predictive control (MPC), offer dynamic optimization in real-time without needing complete trip information. MPC, in particular, runs a continuous optimization process for optimal control, relying on accurate predictive models. These models, essential for optimal real-world performance, consider various dynamic factors like battery aging and driver behavior, which are challenging to fully incorporate.

Reinforcement learning (RL) is one of the machine learning algorithms, and the other two ML algorithms are supervised learning and unsupervised learning. Different from the other two, RL is sometimes regarded as a general process of decision-making tasks, in which the agent learns from experiences that involve interacting with the environment rather than extensive sample data like the other two learning algorithms. Therefore, RL is a feasible technique that has recently been used in various fields, for example, behavioral planning is conducted to identify efficient and satisfied behavior strategies in autonomous driving, and smart energy management is optimized to achieve energy-saving and fuel reduction in smart grids, building, and electric vehicles, etc. Over the past few years, deep reinforcement learning (DRL) has proven to be an effective path for a variety of artificial intelligence problems, which were developed in response to the success of deep neural networks (DNN) at solving complex issues in the real world. Since some games, like the Atari console game, gain profound success through deep Q-network algorithms without complicated tuning in network architecture and hyperparameters, most DRL techniques in the single-agent scenario have nowadays undergone much research and significant advancement [17].

CHEVs in mobility and traffic flow include a variety of agents working together to find a common solution when multiple agents are involved in many real-world intelligent mobility and transport scenarios, which offers an obvious opportunity for multi-agent reinforcement learning (MARL) methods by improving how agents interact and collaborate to achieve the multi-agent system control [11,18]. For example, a multi-agent deep reinforcement learning (MADRL) approach is proposed with all the agents receiving centralized training to create coordinated control strategies for electric vehicles, and make decisions based on the local information when the training process is complete, the long-short-term memory network is employed to predict the uncertainties caused by the load demand and electricity price. MADRL is essential and attractive although it is fundamentally more challenging and complex than the single agent, owing to the increase of dimensionality, multi-agent credit arrangement, and the communication mechanism. For example, there are few entirely homogeneous agents with the same individual attributes and clock synchronization capability in real-world multi-agent applications and uncertain heterogeneous multi-agent systems for vehicle-to-everything (V2X) communications. A MARL method combined with an advanced policy gradient algorithm to investigate cooperative eco-driving system control for CHEVs has been proposed by considering random arrivals and incomplete observation of the environment [19].

Significant advancements in single-agent RL and MARL have been made across different aspects of HEVs and CHEVs on simulation platforms. However, the differences between these simulations and real-world conditions can sometimes result in decreased effectiveness of the developed policies when applied to real-world situations. Due to challenges in collecting real-world data, such as inefficiency in data samples and high costs, training agents in simulated environments has become essential [20]. Therefore, a close match between the data used in training and their real-world operations will be explored to facilitate the implementation of agents.

The literature is congested with extensive and detailed reviews on the RL-based EMS of HEVs and the MARL algorithms, but fewer researchers have conducted a comprehensive review RL-based from the individual vehicle to multiple vehicles, as shown in Figure 1; as there has been an increasing interest in learning approaches more recently to solve numerous issues for more complicated traffic scenarios, like traffic management, road safety, and autonomous driving. Thus, the key contributions of this review are as follows:

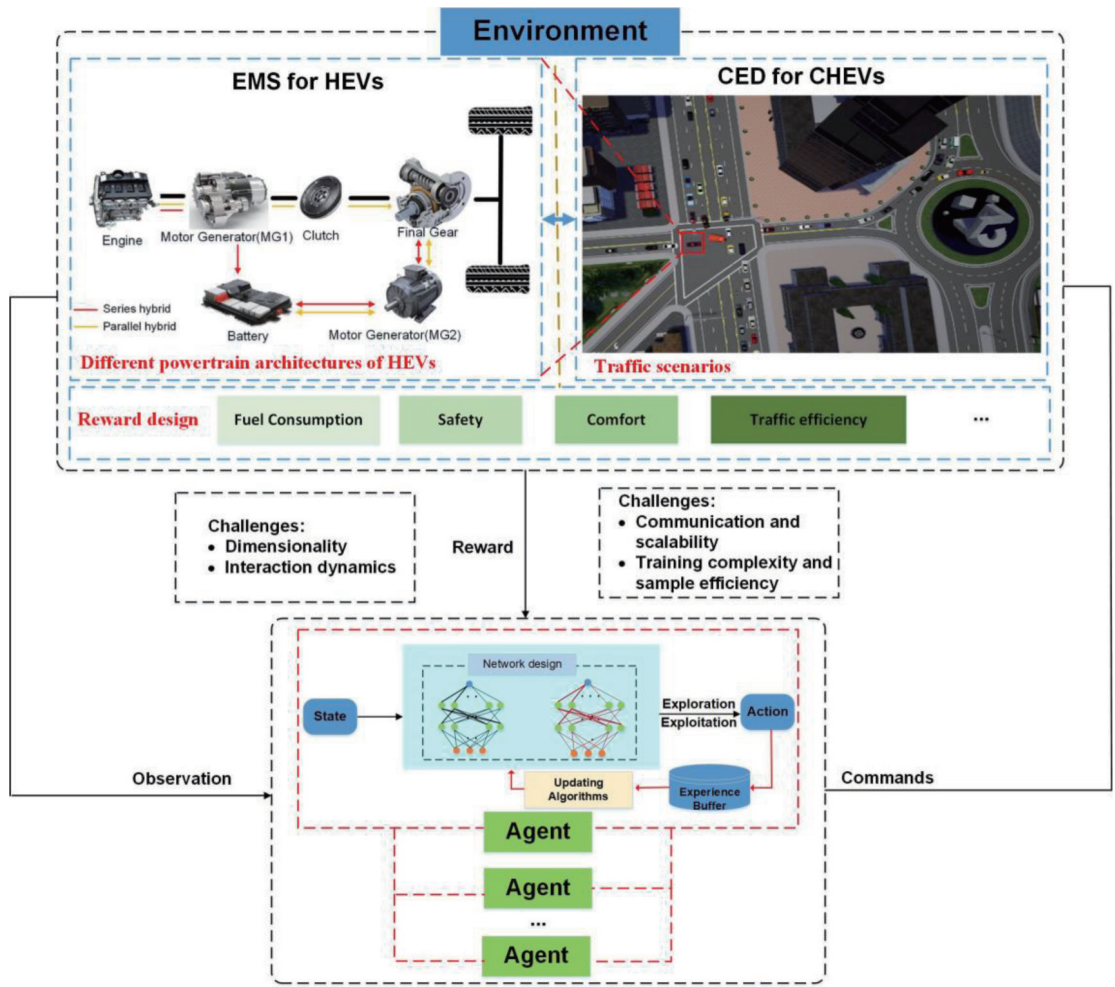

**Figure 1.** Control framework of reinforcement learning (RL) applied in energy management of hybrid electric vehicles (HEVs) and connected hybrid electric vehicles (CHEVs).

(1) Despite RL's promise, a notable gap exists in the current research landscape regarding a comprehensive review of RL-based strategies, particularly across various vehicle scenarios. The absence of such a comprehensive overview hinders a holistic understanding of RL's applicability and potential within CHEVs.

(2) This review paper aims to bridge the knowledge gap by conducting an exhaustive analysis of RL-based solutions. By encompassing a broad spectrum ranging from individual vehicles to multivehicle configurations, the review delves into the challenges, advancements, and untapped possibilities inherent in RL-based approaches.

(3) The evolution of EMS, transitioning from HEVs to CHEVs, marks a pivotal shift. The paper contributes to the foundational knowledge base that will shape the trajectory of sustainable transportation systems in the future.

The remainder of this paper is organized as follows: Section 2 gives a brief summary of RL algorithms in distinguished areas, and then a detailed review of single-agent RL-based energy management systems is described in Section 3. Section 4 summarizes the MARL-based eco-driving management for CHEVs; Section 5 discusses the current challenges and several feasible solutions.

## 2. Reinforcement Learning (RL)

In single-agent RL, an agent interacts with an environment, making a sequence of decisions and actions over time, while receiving feedback in the form of rewards or penalties based on the outcomes of its actions. In MARL, agents have individual objectives and must coordinate their actions. Thus, the intersection of control theory, optimization, and network theory presents a fascinating and complex landscape [19]. Control theory is vital in managing the dynamics and interactions between different agents, ensuring that each agent responds appropriately to both its environment and the actions of other agents. This is crucial to maintaining system stability and achieving desired collective behaviors. Optimization plays a key role in refining these interactions and decisions, aiming to achieve the best possible outcome based on predefined criteria such as fuel efficiency, travel cost, or driving comfort. It involves finding the most effective strategies for each agent and considering the overall objectives of the system. Network theory provides the backbone for understanding and enhancing communication and interconnectivity among agents. It focuses on how information is exchanged, how agents influence each other, and how network structures impact the overall system performance. By integrating these three disciplines, we can develop multi-agent systems that are more efficient, robust, and adaptive, capable of handling complex tasks and environments with greater coordination and intelligence. Research manuscripts reporting large datasets that are deposited in a publicly available database should specify where the data have been deposited and provide the relevant accession numbers. If the accession numbers have not yet been obtained at the time of submission, please state that they will be provided during review. They must be provided prior to publication. Interventional studies involving animals or humans, and other studies that require ethical approval, must list the authority that provided approval and the corresponding ethical approval code.

### 2.1. Single-Agent Reinforcement Learning

In single-agent RL, the objective of the agent is to find the optimal action trajectory by maximizing the expected cumulative rewards [21]. Q-learning and State-Action-Reward-State-Action (SARSA) are both fundamental RL methods based on Q-table, to address the curse of dimension, deep Q-network (DQN) is presented by a combination with Q-learning algorithm and deep learning with SGD training in parallel. For environments with both a continuous action and observation space, Deep Deterministic Policy Gradient (DDPG) is the most straightforward compatible algorithm [22], followed by Twin-Delayed Deep Deterministic Policy Gradient (TD3) [17], Proximal Policy Optimization (PPO), and Soft Actor-Critic (SAC). For such environments, TD3 is an improved, more complex version of DDPG; and PPO has more stable updates but requires more training. SAC is an enhanced, more complex version of DDPG that generates stochastic policies [23]. The learning process of single-agent RL is the Markov Decision Process (MDP) as follows:

(1) The agent can perceive the state $s \in S$ ($S \rightarrow$ StateSpace) of the external environment and the reward $r_t$, and then conduct learning and decision-making at the timestep $t$.

(2) The action $a \in A$ ($A \rightarrow$ ActionSpace) is the behavior of the agent, which can be discrete or continuous, and its action space refers to all actions taken by the agent at this state. The policy $\pi$ is a function of the agent to decide the next action according to the environment. Random strategies $p\ [s_{t+1}|s_t, a_t]$ are generally used due to the better performance of exploring the environment, which is the action-taken probability from the state $s_t$ to next state $s_{t+1}$.

(3) For the environment outside the agent, the environment changes its state under the agent's action and then feeds back the corresponding rewards to the agent. The state is the representation of the environment which can be discrete or continuous, and its state space is all the possible states. The state transition of the environment to the next state occurs, and the immediate reward is obtained after the agent makes an action according to the current state. Ultimately, the cumulative rewards with discount factor $\gamma$ from the timestep $t$ to the end of the timestep $T$ are calculated by:

$$R_t = \sum_t^T \gamma^{t-T} r_t \tag{1}$$

Then the value function $Q_\pi(s,a)$ is the expectation value of the cumulative rewards at the action a and the state $s$ until this episode ends as described by:

$$Q_\pi(s,a) = E_\pi[R_t \,|\, s_t = s, a_t = a] \tag{2}$$

According to the Bellman optimality equation, the optimal state-action value function obtained by an optimal policy $\pi_*$ in all possible policies for all states is as follows:

$$Q_*(s,a) = E_{\pi_*}[r + \gamma \max_{a'} Q(s', a') \,|\, s, a] \tag{3}$$

Therefore, a commonly used approach in RL is the value-based method, which involves estimating the state-action value function. This estimation is accomplished using parameterized function approximators such as neural networks. On the other hand, there is the policy-based method that utilizes the policy gradient to update the network parameters based on the cumulative rewards [24].

## 2.2. Multi-Agent Reinforcement Learning

Learning in a multi-agent environment poses additional challenges due to simultaneous interactions between agents and the environment, as depicted in Figure 2. While all agents collectively influence the environment, each agent individually strives to maximize its cumulative rewards. Consequently, when designing an optimal joint policy, the impacts of joint actions must be carefully considered.

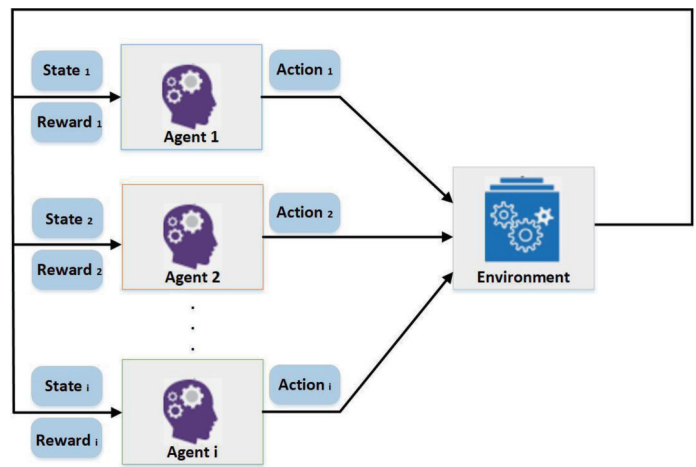

**Figure 2.** Multi-agent reinforcement learning (MARL) framework.

To further understand the learning process of MARL, we consider an MDP game ($S$, $N$, $A$, $P$, $R$) with multiple agents, different from the single-agent RL, the transition $P$, and reward $R$, depend on the action $A = A_1 \times \cdots \times A_N$ ($N$ is the number of agents), thus, $R = R_1 \times \cdots \times R_N$ and $T = S \times A_1 \times \cdots \times A_N$. For a learning agent $i$, the value function is decided by the action of the agent $i$ and the remaining agents $-i$ denoted by $-i = N\backslash i$. Thus, the joint action is $a = (a_i, a_{-i})$ policy $\pi(s,a) = \prod_j \pi_j(s, a_j)$:

$$V_i^\pi(s) = \sum_{a \in A} \pi(s,a) \sum_{s \in S} P\left(s, a_i, a_{-i}, s'\right) [R_i\left(s, a_i, a_{-i}, s'\right) + \gamma V_i(s')] \tag{4}$$

Therefore, the optimal policy is described as follows:

$$\pi_i^*\left(s, a_i, \pi_{-i}\right) = \underset{\pi_i}{\mathrm{argmax}} \ V_i^{(\pi,\pi_{-i})}(s) =$$

$$\underset{\pi_i}{\mathrm{argmax}} \sum_{a \in A} \pi_i(s,a_i) \pi_{-i}(s, a_{-i}) \sum_{s \in S} P\left(s, a_i, a_{-i}, s'\right) [R_i\left(s, a_i, a_{-i}, s'\right) + \gamma V_i^{(\pi,\pi_{-i})}(s')] \tag{5}$$

In this context, each agent typically follows an independent policy. However, there are also scenarios where multiple agents operate in a coordinated or mixed setting. Introducing various settings and combinations can give rise to non-stationary problems, the curse of dimensionality, and challenges in credit assignment over time. Furthermore, there are several intriguing and promising areas that have emerged within the realm of complex multi-agent systems. These areas encompass game theory, cooperative scenarios, evolutionary dynamics, learning in non-stationary environments, agent modeling, and transfer learning in multi-agent RL.

### 2.3. Algorithm Frameworks for Multi-Agent Reinforcement Learning (MARL)

Different structures are employed and studied in multi-agent environments with centralized and decentralized methods. According to different learning and execution modes, there are three main multi-agent frameworks among vast publication and open-source libraries, including decentralizing training and decentralizing execution (DTDE), centralized training and centralized execution (CTCE), and centralized training and decentralized execution (CTDE) [25–27].

#### 2.3.1. Decentralized Learning and Decentralized Execution (DLDE)

In decentralized learning and decentralized execution (DLDE), as shown in Figure 3a, the learning phase involves each agent learning its optimal policy by interacting with the environment and taking actions that maximize its cumulative rewards. This learning is typically done using various reinforcement learning algorithms tailored for multiagent scenarios, such as Q-learning, Policy Gradient methods, or Actor-Critic architectures. After the learning phase, during the execution phase, each agent applies the policy it has learned to make decisions in real time without requiring communication or coordination with other agents. This decentralized execution approach allows agents to act autonomously and quickly, making decisions based on their individual experiences and learned policies.

Common algorithms used in this context include independent Q-learning (IQL), independent advantage actor-critic (IA2C), and independent proximal policy optimization (IPPO), which are often used as baselines. These algorithms allow agents to learn independently but may not capture the full potential of collaboration [28]. DLDE is particularly useful in scenarios where communication between agents is limited, costly, or unreliable. It also provides a scalable solution for environments with a large number of agents, as each agent maintains its own learned policy rather than having to coordinate with all other agents during execution. However, it's important to note that DLDE might not always yield globally optimal solutions, as the agents are acting independently and might not take into account the potential impact of their actions on other agents.

#### 2.3.2. Centralized Learning and Decentralized Execution (CLDE)

In centralized learning and decentralized execution (CLDE), during the learning phase, agents collaborate to optimize a centralized policy that takes into account the actions of all agents. This centralized policy can capture global information and dependencies between agents, enabling more effective cooperation and avoiding conflicts, whereas a centralized critic network estimates the value function based on the actions of all agents. After learning the centralized policy, during the execution phase, agents act independently based on their individual learned policies, without requiring real-time communication with others. This decentralized execution allows agents to make decisions swiftly and efficiently while still benefiting from the collaboration and coordination established during the learning phase, as shown in Figure 3b.

CLDE can be advantageous in scenarios where agents need to collaborate to achieve collective objectives but also need to operate independently in real-world environments where communication might be limited or delayed. It strikes a balance between centralized planning and decentralized execution, offering the benefits of cooperation during learning while maintaining the efficiency of individual decision-making during execution.

### 2.3.3. Centralized Learning and Centralized Execution (CLCE)

In CLCE, during the learning phase, agents work together to learn a single centralized policy that takes into account the actions and observations of all agents. This centralized policy allows for more effective collaboration and coordination, as agents can explicitly learn how their actions impact the global system's performance and goals. Techniques like centralized training with shared experience replay or centralized critics are often used to facilitate the learning of the centralized policy. After learning the centralized policy, during the execution phase, agents continue to act in a coordinated manner based on the shared centralized policy. This centralized execution ensures that agents make decisions that are aligned with the collective goals and knowledge acquired during training. This approach is particularly beneficial in scenarios where precise coordination and cooperation among agents are crucial for achieving optimal outcomes, as shown in Figure 3c.

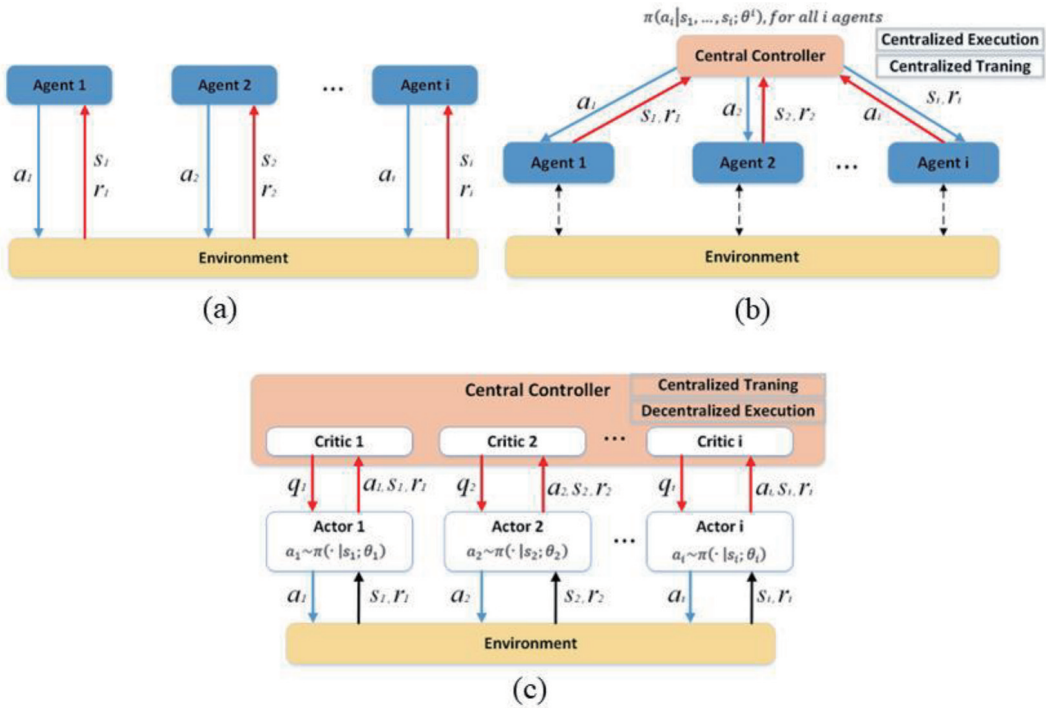

**Figure 3.** Three mainstream frameworks for MARL: (**a**) Decentralized learning and decentralized execution (DLDE); (**b**) Centralized learning and decentralized execution (CLDE); (**c**) Centralized learning and centralized execution (CLCE).

CLCE is suitable for applications where the cost of communication between agents during execution is low, or where real-time coordination is essential for successful task completion. It can be especially effective in situations where the actions of different agents are highly interdependent and require synchronized decision-making [29].

## 3. Single-Agent RL-Based Energy Management System

### 3.1. Classification of Hybrid Electric Vehicles (HEVs)

Hybrid electric vehicles (HEVs) play a significant role in the landscape of transportation electrification, prompting substantial investments in research for the advancement of specialized HEV technologies [30,31].

The measure of hybridization pertains to the proportion of power contributed by an electric motor within a hybrid vehicle relative to the total power utilized by the vehicle. This spectrum of hybridization levels encompasses micro, mild, and full HEVs, each characterized by distinct ratios of electric-to-total power consumption, as illustrated in Figure 4. It is a pyramid-shaped diagram with three distinct sections, each representing different levels of hybrid vehicle technology based on the interaction between the engine, the electric motor, and the battery. At the top of the pyramid, there are two sections labeled parallel and series. In a parallel hybrid system, both the engine and the electric motor are connected in a parallel configuration to the transmission system. They can both directly drive the vehicle either separately or together. This section of the pyramid is subdivided into three levels indicating increasing levels of hybridization: micro hybrid, mild hybrid, and full hybrid. As we move up, the electric motor's role in propelling the vehicle increases. Then, in a series hybrid system, the engine is not directly connected to the wheels but instead is used to power a generator. The generator, in turn, either charges the battery or provides power to the electric motor, which drives the vehicle. Similar to the parallel section, it has incremental stages of hybridization from range extender to series hybrid and finally full hybrid, with the latter having the largest electric motor and battery capacity, allowing for longer electric-only driving capability.

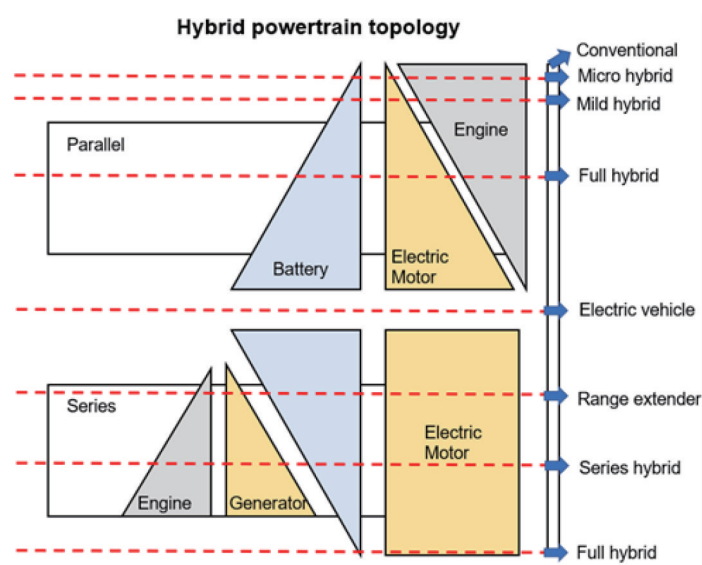

**Figure 4.** Classification of HEVs according to the hybridization degree.

### 3.2. The Application of RL in HEVs

The automotive industry is witnessing a significant trend towards vehicle networking and automation, powered by artificial intelligence. Recent advancements in technologies such as machine vision and deep learning have expedited the progress of advanced assisted driving and energy management control. This has led to extensive research focusing on vehicle perception, intelligent driving, and interactions between drivers, vehicles, and environments. Notably, AI algorithms employing learning-based strategies, including neural-network-based and reinforcement-learning-based approaches, have emerged as research focal points within EMS [32,33].

Neural-network-based strategies exhibit proficiency in handling nonlinear modeling, classification prediction, speed prediction, parameter optimization, and working condition classification. In contrast, the RL algorithm draws inspiration from behaviorist psychology. Agents explore to optimize control strategies, interact with the external environment, update strategies based on feedback, and maximize cumulative rewards in a model-free manner. RL-based algorithms have seen notable development in EMS research for various electrified vehicles, such as hybrid electric vehicles, plug-in hybrid electric vehicles, and battery/fuel cell/ultra-capacitor vehicles, as evident in Table 1 [34].

**Table 1.** Single-agent reinforcement learning (RL) -based energy management systems (EMS) for hybrid electric vehicles (HEVs).

| Author's | Configuration | RL Algorithm | Action | State | Reward |
|---|---|---|---|---|---|
| Xu et al. [38] | HEVs (parallel) | Q-learning | Power split | Velocity, Power demand, Acceleration, SoC | Fuel Consumption |
| Liessner et al. [39] | HEVs | Deep Q Network (DQN) | Power output for electric motor | Wheel speed, SoC | Fuel Consumption |
| Liu et al. [40] | HEVs, PHEVs | Q-learning, Dyna H | Power output for electric motor | Velocity, SoC | Fuel Consumption |
| Hu et al. [41] | HEVs | Deep Q Network (DQN) | Engine Torque | Total required torque | Fuel Consumption |
| Zhou et al. [42] | FCEVs | Deep DDPG | Upper and lower limits of SoC | SoC | Fuel Consumption |
| Song et al. [43] | PHEVS (parallel) | Deep Q Network (DQN) | Engine Torque | SoC, Power demand, Velocity | Fuel Consumption |
| Hsu et al. | Fuel Cell HEVs | Q-learning (Tabular) | Degree of Hybridization | SoC, Pedal position | Fuel Consumption |
| Sun et al. [44] | Fuel Cell HEVs | Q-learning (Tabular) | Battery Power | SoC | Fuel Consumption |
| Reddy et al. [45] | Fuel Cell HEVs | Q-learning (Tabular) | Change in fuel cell power | SoC | Fuel Consumption |
| Wu et al. [46] | HEVs | Deep Q Network (DQN) | Power of internal combustion engine | SoC, Power demand, Velocity | Fuel Consumption |
| Han et al. [47] | HEVs | Double DQN | Rotational speed | SoC, Power demand, Velocity | Fuel Consumption |

For different driving styles and changes in common driving routes, controllers can adaptively finetune control strategies through online learning to enhance energy efficiency and reduce emissions. RL-based EMS enables onboard optimization. Among the array of RL algorithms discussed, the Q-learning algorithm employing a tabular method holds prominence, followed by the Deep Q Network. This preference is attributed to computational efficiency and algorithmic complexity. While the Q tabular method demands more memory computation, it updates significantly faster than NN-based RL, making it more suitable for practical driving cycles. To this end, this study designs an EMS controller for a plug-in hybrid electric vehicle (PHEV) based on tabular Q learning algorithms, showcasing their superior performance in PHEV optimization [35–37].

Through analyzing the basic principle of RL, RL can be employed in the EMS for the HEVs. The vehicle information (e.g. battery SoC, driver's power demand $P_{dem}$) controller represents an agent, which directly takes actions (e.g. control command a) and makes interactions with the unknown driving environments (e.g. road, driver behavior, surrounding vehicles, etc.). The tasks of RL for the EMS of the HEV are usually described by using the MDP, as shown in Figure 5. According to the state observation, such as the power demand $P_{dem}$, the agent (EMS unit) takes the action to HEV-based maximum value policy, and the reward (considering the fuel consumption from HEVs) will be obtained.

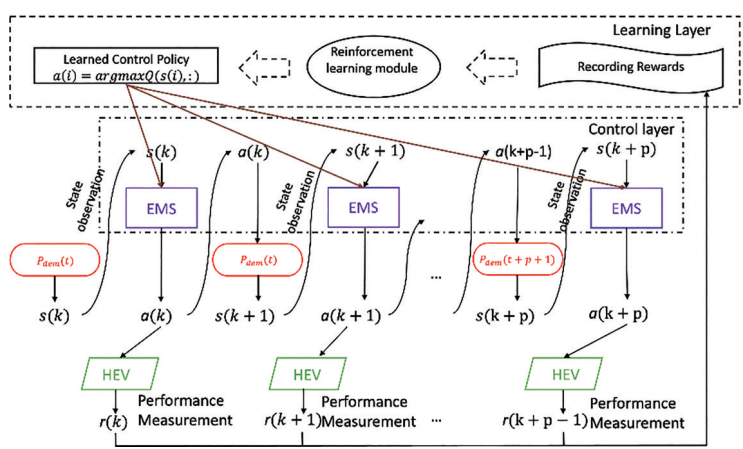

**Figure 5.** RL applied to energy management systems (EMS) of EVs with Markov Decision Process (MDP) process.

## 4. Multi-Agent RL-Based Cooperative Eco-Driving System

### 4.1. Typical Traffic Scenarios

CHEVs encounter a range of intricate traffic scenarios, each highlighting their adaptability and transformative potential in reshaping transportation [48,49].

As shown in Figure 6, platooning means CHEVs drive close together to save fuel. Trucks and big vehicles form groups, with a leader setting the speed and guiding others. This reduces air resistance, helping all vehicles use less fuel and emit fewer gases. Automation keeps them in sync, making roads safer and traffic smoother. Even though platooning has challenges like communication and rules, it's a great way to make long-distance travel efficient and eco-friendly. Wang et al. describe a CED system targeted for signalized corridors, focusing on how the penetration rate of CHEVs affects the energy efficiency of the traffic network and propose a role transition protocol for CHEVs to switch between a leader and following vehicles in a string [50].

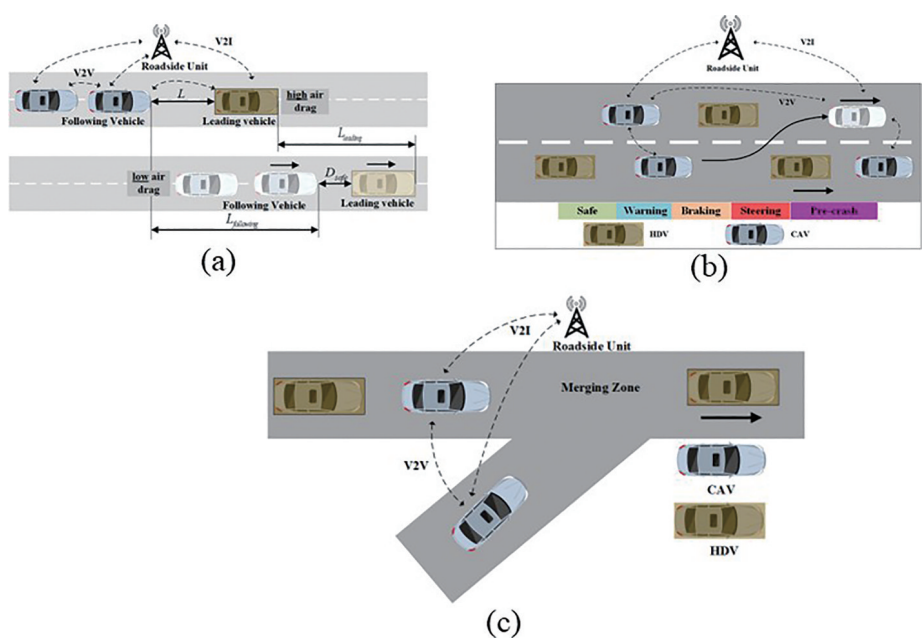

**Figure 6.** Different traffic scenarios: (**a**) Platooning; (**b**) Lane-changing scenario; (**c**) Merging on the highway.

Moving to two-dimensional traffic scenarios, lane-changing of CHEVs within an eco-driving context involves automated maneuvers aimed at optimizing fuel efficiency and minimizing emissions while transitioning between lanes. CHEVs equipped with advanced sensors, communication systems, and eco-driving algorithms assess real-time traffic conditions and environmental factors to determine the most energy-efficient moments for lane changes. By harmonizing lane-changing decisions with eco-driving strategies, CHEVs contribute to a more sustainable transportation ecosystem while maintaining safety and overall road efficiency. In [51], the longitudinal control controlled by DRL, and the lateral control designed by DQN to solve the CED control have been conducted.

In more complex situations, for instance, when CHEVs need to merge into traffic or follow traffic lights, they use tech to optimize energy efficiency, safety, and traffic flow. When merging, CHEVs use sensors to join traffic smoothly, saving energy and helping traffic move better. Traffic light control involves CHEVs communicating with traffic infrastructure to anticipate signal changes and adjust speeds accordingly, reducing the need for sudden stops and accelerations. This integrated approach ensures not only reduced fuel consumption and emissions but also enhanced traffic coordination, making substantial strides toward an eco-friendly and efficient transportation ecosystem. Liu et al. propose an intermediate solution to improve the traffic efficiency of CHEVs in unsignalized intersections by avoiding congestion through coordinated behavior controlled by a deep reinforcement learning agent with an altruistic reward function [52].

*4.2. The Application of Multi-Agent RL-Based Cooperative Eco-Driving*

The objective of an energy management system (EMS) is to improve the internal energy flow of electric vehicles (HEVs). The dynamics of the ego vehicle are typically used to define the state. However, the usage of CHEVs in transportation is expected to increase. With the development of 5G cellular communication technology and autonomous driving, CHEVs can interact with each other in real-time, allowing for cooperative driving behavior [53,54]. Typically, the CED control aims to leverage this capability to optimize the driving behavior of multiple CHEVs in a way that minimizes energy consumption, improves traffic flow, and enhances driving comfort. Moreover, the CED control for CHEVs is also a promising solution for addressing safety concerns. By enabling CHEVs to communicate with each other, they can share information about their surroundings, such as obstacles, road conditions, and traffic flow, allowing them to make more informed decisions and avoid accidents. Additionally, by optimizing the driving behavior of multiple CHEVs, the likelihood of collisions and sudden braking can be reduced, improving the overall safety of the transportation system. Therefore, cooperative control for CHEVs is an essential research area in the development of sustainable and efficient transportation systems [55].

More specially, the CED control considers an explicit relationship between ego vehicle dynamics (e.g., longitudinal acceleration/deceleration) and energy consumption and aims to optimize ego vehicle dynamics given environmental information (e.g., surrounding vehicles, signals, etc.). In terms of CED, the ego vehicle is usually regarded as the agent, and there are typically one or more CHEVs as ego vehicles in one scenario; the longitudinal (i. e., acceleration and deceleration) and lateral (i.e., lane changing) operations of the ego vehicle can be modeled as actions. Driving scenarios, including dynamics of surrounding vehicles and the physical structure of intersections and various lanes, can be considered as the environment. Then fuel consumption and travel time can be used as rewards.

Conventional model-based eco-driving approaches typically necessitate sophisticated models and are incapable of dealing with complex driving scenarios. With the successful applications of artificial intelligence in some fields, such as robotics, game theories, and autonomous driving, a variety of multiagent problems have been presented and solved to obtain some excellent performances [56]. To model the multi-agent system, the most straightforward MARL method is independent RL (IRL), IRL is scalable from the perspective of any individual agent because all agents are completely independent without information exchange [57]. Guo et al. propose a hybrid deep reinforcement learning approach to optimize the energy efficiency of CHEVs by utilizing real-time information from sensors, cameras, and traffic lights. The deep deterministic policy gradient method is designed for longitudinal control and deep Q-learning is designed for lateral control to solve the eco-driving model [51]. Therefore, it suffers from instability, and all other agents' policies are implicitly formulated as part of the environment dynamics while their policies are continuously adjusted during the training process. To solve this issue, a multi-agent deep deterministic policy gradient (MADDPG) method is created that incorporates all agents' input into the training process [58]. Zhou et al. propose a MADDPG-based approach for cooperative lane change maneuvers of ACVs. The proposed method utilizes MADDPG to learn a joint policy for a group of CHEVs to coordinate their lane change maneuvers and improve traffic flow. Results showed that the proposed method outperforms existing methods in terms of safety and efficiency [42]. Moreover, one of the limitations of MADDPG in CHEVs is its high computational complexity and training time. As CHEVs require real-time decision-making and continuous learning, reducing the computational cost of MADDPG is crucial for its practical implementation. To address this challenge, some studies have proposed hybrid approaches that combine MADDPG with other algorithms to improve its efficiency and scalability [59]. For instance, Liu et al. propose a hybrid algorithm that combines MADDPG with a rule-based method to reduce the computational cost of the joint policy learning process.

MARL can be categorized into four main classifications based on their methods of communication. The initial category involves non-communicative approaches, primarily enhancing training stability through advanced value estimation techniques. In MADDPG, a centralized critic uses global observations and actions to estimate each action value. Counterfactual multi-agent (COMA) extends this concept to A2C, estimating advantages using a centralized critic and a counterfactual baseline [60]. In Dec-Hysteretic deep recurrent Q-Networks (HDRQN), the centralized critic employs local observations with global parameter sharing [61]. Zhang et al.'s NMARL work employ fully decentralized critics, each working with global observations and consensus updates. In this study, the effectiveness of introducing a spatial discount factor has empirically been validated to enhance the training stability of non-communicative algorithms when utilizing neighborhood observations. The second group explores heuristic communication protocols and direct information sharing. Foerster et al. demonstrate performance improvements through the direct exchange of low-dimensional policy fingerprints with other agents. Similarly, mean-field MARL computes the average of neighbor policies for informed action-value estimation. This group's notable drawback is the lack of explicit design for communication optimization, potentially leading to inefficient and redundant exchanges during execution, in contrast to NeurComm. The third one introduces learnable communication protocols. In DIAL, each DQN agent generates a message alongside action-value estimation, which is then encoded and combined with other input signals at the receiving end. CommNet is a more generalized communication protocol, calculating the mean of all messages instead of encoding them. Both methods, particularly CommNet, suffer from information loss due to aggregation of input signals. Communication within strategy games has been concentrated. Bidirectionally-Coordinated Network (BiCNet) employs a bidirectional RNN to enable direct communication among agents, while master-slave employs two-way message passing within a hierarchical RNN structure of master and slave agents. NeurComm, in contrast to existing protocols, employs signal encoding and concatenation to minimize information loss and incorporates policy fingerprints in communication to address non-stationarity. The fourth group emphasizes communication attention for selective message transmission. An attentional communication model learns soft attention, assigning communication probabilities to other agents [62]. On the other hand, IC3Net employs hard binary attention to decide whether to communicate or not. These approaches are particularly valuable when agents need to prioritize communication targets, a scenario less likely in NMARL due to the constrained communication range within small neighborhoods.

## 5. Challenges and Potential Solutions

The increasing adoption of HEVs offers a significant opportunity to revolutionize transportation energy systems. The shift towards electrifying transportation addresses environmental concerns tied to fossil fuel usage, necessitating effective EMS and cooperative eco-driving control for optimizing energy efficiency. The current challenges and the corresponding potential solutions for RL-based energy management spanning from single vehicles to multiple vehicles encompass several critical aspects:

(1) **Dimensionality reduction:** As the transition is made from single vehicles to multiple vehicles, the dimensional complexity of the problem increases significantly. Multi-agent systems involve a larger state and action space, making RL algorithms computationally intensive and prone to slow convergence. Handling this complexity while maintaining real-time decision-making poses a considerable challenge. Therefore, dimensionality reduction techniques will be employed, such as principal component analysis (PCA), autoencoders, and linear discriminant analysis (LDA), to simplify state and action spaces without compromising critical information.

(2) **Interaction dynamics:** In the context of multiple vehicles, intricate interaction dynamics arise due to cooperative control, communication, and potential conflicts. Coordinating decisions among different agents becomes challenging, as optimal actions for one vehicle may impact others' behaviors and overall system performance. Developing effective RL algorithms that balance individual and collective objectives is a complex endeavor. Therefore, employing techniques such as hierarchical RL or game theory to model interactions and incentives among agents will be conducted, fostering effective cooperation and preventing conflicts. Further, decentralized RL approaches that enable agents to make decisions based on local observations while optimizing global objectives will be implemented. Techniques like decentralized policy optimization or message-passing algorithms will be leveraged to achieve effective coordination without central control.

(3) **Communication and scalability:** Communication and information sharing play a pivotal role in multi-vehicle scenarios in terms of network connectivity in CHEVs. Vehicles exchange information for CED control, posing challenges in terms of communication delays, reliability, and handling communication failures. Additionally, scalability becomes an issue as the number of vehicles increases, as it can strain communication networks and lead to computational inefficiencies through network connectivity and data exchange. Therefore, robust communication protocols that handle delays and failures effectively will be integrated. Communication-efficient RL methods that prioritize essential information exchange and multi-level communication networks to accommodate scalability while minimizing congestion may be utilized.

(4) **Training complexity and sample efficiency:** Training RL-based EMS for multiple vehicles requires substantial training data to capture the complexity of the environment and interactions. Achieving sample efficiency and reducing the number of interactions required for training while ensuring convergence to optimal policies is a significant challenge. Therefore, several techniques for enhanced sample efficiency, such as using off-policy methods or experience replay will be introduced to optimize learning from limited interactions. Curriculum learning or transfer learning to leverage knowledge gained from similar scenarios for faster convergence will be implemented. Addressing these challenges requires a multi-faceted approach that combines techniques from machine learning, multi-agent systems, and control theory. As research continues to evolve, the collaborative effort of experts in these fields will be crucial to creating effective solutions for RL-based energy management in multi-vehicle scenarios.

## 6. Conclusions

In summary, the rise of HEVs is transforming transportation with a focus on energy efficiency. EMS play a crucial role, evolving from traditional vehicles to CHEVs. This shift demands advanced algorithms for HEVs to optimize energy use, charging, and distribution. RL, especially MARL, is a key tool for handling complex scenarios. However, there's a need for a comprehensive review spanning individual vehicles to multi-vehicle setups. This review fills that gap, shedding light on challenges, advancements, and possibilities within RL-based solutions for sustainable transportation systems. Overcoming challenges like dimensionality, interactions, and decision-making requires robust communication, scalable algorithms, and fair reward systems. Personalized rewards, adaptable frameworks, and efficient training methods are promising solutions. Collaboration across fields will lead to innovative RL-based EMS solutions, shaping the future of eco-friendly transportation.

**Author Contributions:** Conceptualization, M.H. and B.S.; methodology, M.H. and Q.Z.; formal analysis, Q.Z., J.W. and Y.H.; investigation, Q.Z.; resources, M.H. and B.S.; data curation, Q.Z. and Y.H.; writing—original draft preparation, M. H.; writing—review and editing, B.S. and Q.Z.; visualization, M.H. and J.W.; supervision, H.X.; project administration, B. S., Q.Z. and H.X.; funding acquisition, Q.Z. and H.X. All authors have read and agreed to the published version of the manuscript.

**Funding:** This research is sponsored by the State Key Laboratory of Automotive Safety and Energy, Tsinghua University, grant number No. KF2029.

**Data Availability Statement:** Not applicable.